\documentclass[12pt]{article}
\usepackage{epsfig}
\usepackage{amsmath}
\def\block{\hbox{${\vcenter{\vbox{\hrule height 0.4pt\hbox{\vrule
width 0.4pt
height 6pt \kern 5pt\vrule width 0.4pt}\hrule height 0.4pt }}}$}}
\def\be{\begin{equation}}
\def\ee{\end{equation}}
\def\bea{\begin{eqnarray}}
\def\eea{\end{eqnarray}}

\def\la{\lambda}
\def\lR{{\hbox{{\rm I}\kern-.2em\hbox{\rm R}}}}
\def\IZ{{\hbox{{\rm I}\kern-.2em\hbox{\rm Z}}}}
\def\IC{{\hbox{{\rm I}\kern-.2em\hbox{\rm C}}}}
\def\II{{\hbox{{\rm I}\kern-.2em\hbox{\rm I}}}}

\begin{document}

\begin{titlepage}
\vspace{.5in}
\begin{flushright}
DFTT/12/2001\\
IST/DM/20/01\\
May 2001\\
math.MP/0105015\\
\end{flushright}
\begin{center}
{\Large\bf
Parametrization of the moduli space of flat ${\rm SL}(2,R)$
connections on the torus}\\ \vspace{.4in}
{J.\ E.~N{\sc elson}\footnote{\it email: nelson@to.infn.it}\\
       {\small\it Dipartimento di Fisica Teorica, Universit\`a degli Studi
       di Torino}\\
{\small\it and Istituto Nazionale di Fisica Nucleare, Sezione di Torino}\\
       {\small\it via Pietro Giuria 1, 10125 Torino}\\
       {\small\it Italy}}\\
\vspace{1ex}
{\small and}\\
\vspace{1ex}
{R.\ F.~P{\sc icken}\footnote{\it  email:  picken@math.ist.utl.pt}\\
{\small\it Departamento de Matem\'{a}tica and Centro de
Matem\'{a}tica Aplicada}\\
{\small\it  Instituto Superior T\'{e}cnico}\\
{\small\it Avenida Rovisco Pais, 1049-001 Lisboa}\\
{\small\it Portugal}}
\\\vglue 5mm
\end{center}
\begin{center}
\begin{minipage}{4.8in}
\begin{center}
{\large\bf Abstract}
\end{center}
{\small

The moduli space
of flat ${\rm SL}(2,R)$-connections modulo gauge transformations on the torus
may be described by
ordered pairs of commuting ${\rm SL}(2,R)$ matrices modulo simultaneous
conjugation by ${\rm SL}(2,R)$ matrices. Their spectral properties allow a
classification of the
equivalence classes, and a unique canonical form is given for each of
these. In this way the moduli space becomes explicitly parametrized, and
has a simple structure, resembling that of a cell complex, allowing it to be
depicted. Finally, a Hausdorff
topology based on this classification and parametrization is proposed. }
\end{minipage}

\vspace{.2in}

\vspace{.2in}


\end{center} \end{titlepage} \addtocounter{footnote}{-2}

\section{Introduction \label{sec1}}

Moduli spaces of flat $G$-connections over a Riemann surface $M$ have attracted a
vast amount of attention in the mathematics and physics literature. For
instance they are of interest as the space of solutions of Chern-Simons
theory, and much effort has been devoted to studying their geometry,
both as symplectic stratified spaces \cite{ati:bot,gur:jef:hue:wei,gol},
and from the point of view of
algebraic geometry \cite{hit}. In most cases the group $G$ is chosen to
be compact, and frequently the Riemann surface is taken to be of genus
greater than or equal to $2$.

As shown by Witten \cite{wit}, Chern-Simons theories with certain non-compact
 groups $G$ are relevant for the study of  $2+1$-dimensional gravity.
When the cosmological constant is negative, $G$ is isomorphic to  ${\rm
SL}(2,R)\times {\rm SL}(2,R)$, and the theory effectively splits into two
Chern-Simons theories with group ${\rm SL}(2,R)$. This approach has been a
useful starting point for describing the quantum theory of $2+1$ gravity
\cite{jeanette,car}. Our own interest in the moduli space of
flat ${\rm SL}(2,R)$ connections on the torus arose precisely from attempts
to understand $2+1$ quantum gravity with negative cosmological constant on
the torus, from a non-local geometry perspective \cite{NP1,NP2}.

Indeed, non-local geometry plays a key role in simplifying the analysis of
the moduli space of smooth flat $G$-connections on $M$ modulo smooth gauge
transformations, an infinite-dimensional space divided by the action of an
infinite-dimensional group. It is well-known that this space may be
identified with ${\rm Hom}(\pi_1(M),G)/G$, where $G$ acts by conjugation,
by using the holonomy of the connections. This
fact may be regarded as a special case of the main result in
\cite{CP}, following earlier work by Barrett \cite{bar}, which makes
precise the correspondence between smooth connections,
not-necessarily-flat, and ``holonomy assignments'' obeying a suitable
smoothness condition. The reduction to ${\rm Hom}(\pi_1(M),G)/G$, in the
case of flat connections, gives a finite-dimensional perspective on the
moduli space whose importance has been emphasized by Huebschmann
in several mathscinet reviews.

The moduli space considered here for a manifold
of genus one and group ${\rm SL}(2,R)$ is closely related to the Teichm\"{u}ller
space of the torus, using the Goldman \cite{gol} description of Teichm\"{u}ller
spaces of (higher genus) surfaces, as ${\rm Hom}(\pi_1(M), G)/G$ with
$G={\rm PSL}(2,R)$, the projective special linear group.

The purpose of this communication is to show that for $G= {\rm
SL}(2,R)$ and $M$ of genus $1$, the non-local geometry viewpoint leads to
a completely explicit description of the moduli space by
 using only elementary tools of linear algebra.
This is appealing, since moduli spaces tend to be complicated
spaces, requiring sophisticated tools, e.g. of algebraic geometry, for
their description. The main observation is that
 $\pi_1$ of the torus ${\bf T}^2$ is the free abelian group on two
generators, and therefore a homomorphism from $\pi_1({\bf T}^2)$ to $ {\rm
SL}(2,R)$  is given by an ordered pair $(U_1, U_2)$ of commuting $ {\rm
SL}(2,R)$ elements, being the images of the two generators under the
homomorphism. That the matrices commute imposes restrictions on
the spectral properties of the matrices in each pair, which we then classify.
Further, these pairs are identified up to
simultaneous conjugation by elements of $ {\rm SL}(2,R)$, which allows us
to find a unique canonical form for each equivalence class. These results
are given in the theorem in Section 2. As a consequence we obtain a
full and explicit parametrization of the moduli space, allowing its
structure to be visualized.

Several informal treatments of the moduli space
under discussion, or closely related ones, have appeared in the physics
literature  \cite{eza,dwi:mat:nic}, \cite{lou:mar} ($G = {\rm
ISO}(2,1)$), \cite{lou} ($M$ the Klein bottle), \cite{fal:gaw} ($G= {\rm
SL}(2,C)$).
Our rigorous approach
via the spectrum and canonical forms may also  be adaptable
to other moduli spaces, and also suggests a natural choice of
topology on the moduli space, which we discuss in Section 3. In
contrast with other authors \cite{lou:mar,ash:lew}, who have
proposed a non-Hausdorff topology, this topology is Hausdorff,
essentially since it separates pairs with spectra of
different types. As a final remark, a treatment of a supersymmetric version
of the moduli space was given by Mikovic and one of the authors in
\cite{mik:pic}.

\section{ The moduli space of flat ${\rm SL}(2,R)$-connections on the torus}

As stated in the introduction, flat ${\rm SL}(2,R)$-connections,
modulo gauge transformations, on the torus ${\bf T}^2$ are in one-to-one
correspondence with group homomorphisms from $\pi_1({\bf T}^2)$ to ${\rm SL}(2,R)$,
 modulo conjugation by an element of ${\rm SL}(2,R)$. Geometrically this
conjugation corresponds to gauge transformations in the fibre over the base
point of the fundamental group.

 The fundamental group of the torus
is the free abelian group on two generators, and thus a
homomorphism $\pi_1({\bf
T}^2)\rightarrow {\rm SL}(2,R)$ is specified by two commuting
$SL(2,R)$ matrices, the values of the homomorphism on two generating
cycles of the fundamental group.
(We will deal throughout with the defining $2\times 2$ matrix
representation of  ${\rm SL}(2,R)$, as opposed to the abstract Lie group.)
The conjugation action of ${\rm SL}(2,R)$ on a homomorphism corresponds to
simultaneous conjugation of these two elements by the same element of ${\rm
SL}(2,R)$. Therefore our moduli space $\cal M$ is defined to be

$$ {\cal
M} :=\left\{ (U_1, U_2)\in  {\rm SL}(2,R) \times  {\rm SL}(2,R)| U_1U_2=U_2U_1
\right\}/\sim
$$
 where
 $$ (U_1, U_2)\sim (U_1', U_2')
\Longleftrightarrow \exists S\in {\rm SL}(2,R)\quad  U_i'=S^{-1}U_iS, i=1,2
\label{equivreln} $$

We start by recalling the classification of a single ${\rm SL}(2,R)$ matrix
$U$ in terms of its spectral properties:

\begin{itemize}
\item[A)] $U$ has two real eigenvalues $\lambda$ and
$\lambda^{-1}$;
\item[B)] $U$ has one real eigenvalue $\pm 1$ with an eigenspace of
dimension two;
\item[C)] $U$ has one real eigenvalue  $\pm 1$ with an eigenspace of
dimension one;
 \item[D)] $U$ has
no real eigenvalues.
\end{itemize}
\noindent These cases may be partly distinguished by the trace of $U$: case
A) corresponds to $|{\rm tr}\, U|>2$, cases B) and C) to $|{\rm tr}\,
U|=2$, and case D) to $|{\rm tr}\, U|<2$.

\noindent In case A) $U$ may be conjugated
to diagonal form: $$ \exists S\in {\rm GL}(2,R) \quad S^{-1}U S =
\left(\begin{array}{ll}\lambda & 0 \\ 0 & \lambda^{-1}\end{array}\right).
$$ In case B) $$ U=\pm \left(\begin{array}{ll} 1& 0 \\0&1
\end{array}\right). $$ In case C) $U$ may be conjugated to upper-triangular
Jordan canonical form $$ \exists S\in {\rm GL}(2,R) \quad S^{-1}U S =
\left(\begin{array}{ll}\pm 1 & 1 \\ 0 & \pm 1\end{array}\right). $$ In case
D) $U$ has complex conjugate eigenvalues $e^{\pm i\theta}$, and may be
conjugated to the form of a rotation matrix by a negative angle (real
Jordan canonical form) $$ \exists S\in {\rm GL}(2,R) \quad S^{-1}U S =
\left(\begin{array}{ll}\cos \theta & -\sin \theta \\ \sin \theta & \cos
\theta \end{array}\right), \quad -\pi<\theta< 0. $$

If we introduce an equivalence relation on ${\rm SL}(2,R)$ matrices
$$
U\sim U' \Longleftrightarrow \exists S \in {\rm GL}(2,R) \quad U'=S^{-1}US
$$
then the diagonal or Jordan canonical forms above provide a natural choice
of representative for each equivalence class, which is furthermore unique,
except for the order of the eigenvalues on the diagonal in case A). The
analogous problem to be solved here is to find a natural and unique
canonical form for commuting pairs of ${\rm SL}(2,R)$ matrices up to
simultaneous conjugation by elements of ${\rm SL}(2,R)$. We remark that the
restriction to conjugation by ${\rm SL}(2,R)$ elements instead of ${\rm
GL}(2,R)$ elements has consequences even for a single matrix. For instance
the rotation matrices for angles $\theta$ and $-\theta$ are only conjugate
when using ${\rm GL}(2,R)$ elements, not when using ${\rm SL}(2,R)$
elements (see the proof below).

\pagebreak
\noindent {\bf Theorem} {\em
Let $(U_1,U_2)$ be a pair of commuting ${\rm SL}(2,R)$ matrices. In terms
of the previous spectral classification into types {\rm A)-D)}, the possible
combinations of types for $(U_1,U_2)$ are {\rm (A,A), (C,C), (D,D),
(B,$\ast$) and ($\ast$, B)}, where $\ast$ denotes any type. Under
simultaneous conjugation by $S\in {\rm SL}(2,R)$, any pair may be put
uniquely into one of the following forms: }


\[
\begin{array}{lcl}

{\rm (AA1)} &

\left[
\left( \begin{array}{cc} \la &0\\ 0&\la^{-1}\end{array} \right),
\left( \begin{array}{cc} \mu &0\\ 0&\mu^{-1}\end{array} \right)
\right],  & 0<|\lambda|<1, \quad  0<|\mu|<1, \\

 & & \\

{\rm (AA2)}
 &
\left[
\left( \begin{array}{cc} \la &0\\ 0&\la^{-1}\end{array} \right),
\left( \begin{array}{cc} \mu^{-1} &0\\ 0&\mu \end{array} \right)
\right], &  0<|\lambda|<1, \quad 0<|\mu|<1, \\

 & & \\

{\rm (AB)}
   &
\left[
\left( \begin{array}{cc} \la &0\\ 0&\la^{-1}\end{array} \right),
\left( \begin{array}{cc} \epsilon_2 &0\\ 0 & \epsilon_2 \end{array} \right)
\right], &  0<|\lambda|<1, \quad  \epsilon_2\in \{+1,-1\}, \\

  & & \\

{\rm (BA)}
&
\left[
\left( \begin{array}{cc} \epsilon_1 &0\\ 0&\epsilon_1\end{array} \right),
\left( \begin{array}{cc} \mu &0\\ 0&\mu^{-1} \end{array} \right)
\right], &  \epsilon_1\in \{+1,-1\}, \quad  0<|\mu|<1, \\

  & & \\

{\rm (BB)}
&
\left[
\left( \begin{array}{cc} \epsilon_1 &0\\ 0&\epsilon_1\end{array} \right),
\left( \begin{array}{cc} \epsilon_2 &0\\ 0 & \epsilon_2 \end{array} \right)
\right], &  \epsilon_1, \epsilon_2 \in \{+1,-1\},  \\

 & & \\

{\rm (BC)}
&
\left[
\left( \begin{array}{cc} \epsilon_1 &0\\ 0&\epsilon_1\end{array} \right),
\left( \begin{array}{cc} \epsilon_2 &\epsilon_4 \\ 0 & \epsilon_2
\end{array} \right) \right], &   \epsilon_1, \epsilon_2, \epsilon_4 \in
\{+1,-1\},  \\

 & & \\




{\rm (CB)}
&
\left[
\left( \begin{array}{cc} \epsilon_1 &\epsilon_3 \\ 0 &\epsilon_1\end{array}
\right), \left( \begin{array}{cc} \epsilon_2 &0\\ 0 & \epsilon_2
\end{array} \right) \right], &   \epsilon_1, \epsilon_2, \epsilon_3 \in
\{+1,-1\},
\\

 & & \\

{\rm (BD)}
&
\left[
\left( \begin{array}{cc} \epsilon_1 &0\\ 0&\epsilon_1\end{array} \right),
\left( \begin{array}{cc} \cos \phi & -\sin \phi \\ \sin \phi & \cos \phi
\end{array} \right) \right], &  \epsilon_1 \in \{+1,-1\}, \quad
\phi \in ]0, \pi[ \cup ]\pi, 2\pi[, \\

  & & \\

{\rm (DB)}
&
\left[
\left( \begin{array}{cc}\cos \theta & -\sin \theta \\ \sin \theta & \cos
\theta \end{array} \right), \left( \begin{array}{cc} \epsilon_2 &0\\ 0 &
\epsilon_2 \end{array} \right) \right], & \theta  \in ]0, \pi[ \cup
]\pi, 2\pi[,  \quad \epsilon_2 \in \{+1,-1\},
\\

 & & \\

{\rm (CC)}
&
\left[
\left( \begin{array}{cc} \epsilon_1 &\cos \alpha\\ 0&\epsilon_1\end{array}
\right), \left( \begin{array}{cc} \epsilon_2 &\sin \alpha\\ 0 & \epsilon_2
\end{array} \right) \right], &  \epsilon_1, \epsilon_2 \in \{+1,-1\},
\, \\

 & &   \alpha \in\,  ]0,2\pi\, [\setminus\left\{\frac{\pi}{2}, \pi,
\frac{3\pi}{2}\right\},
  \\

 & & \\

{\rm (DD)}
&
\left[
\left( \begin{array}{cc}\cos \theta & -\sin \theta \\ \sin \theta & \cos
\theta \end{array} \right),
\left( \begin{array}{cc} \cos \phi & -\sin \phi \\ \sin \phi & \cos \phi
\end{array} \right)
\right], & \theta, \phi  \in ]0, \pi[ \cup
]\pi, 2\pi[.

\end{array}
\]


\vskip 1cm

\noindent {\em Proof} a) Since the pairs {\rm (B,$\ast$) and ($\ast$, B)} are
obviously commuting, it is enough to show that no combinations of A), C)
and D) can occur, other than {\rm (A,A), (C,C), (D,D)}. This follows from
the fact that, for commuting matrices, any eigenvectors are joint
eigenvectors, since the number of real $1$-dimensional eigenspaces differs
for the three cases (2, 1 and 0 for A), C), and D) respectively).
\vskip .2cm
\noindent b)
We only consider pairs in alphabetical order, since the
reasoning for the remaining pairs is identical.
\vskip .2cm
\noindent (AA) Let $\{v_1, v_2\}$ be a pair of linearly independent joint
eigenvectors of the two matrices. Thus they are simultaneously diagonalized
by the matrix $S'\in {\rm GL}(2,R) $, whose columns are $v_1$ and $v_2$,
and this diagonal form is unique up to the ordering of the eigenvalues on
the diagonal. Rescaling one of the columns of $S'$ by $1/\det S'$ gives a
matrix $S\in  {\rm SL}(2,R) $, which also diagonalizes both matrices.
Finally the first diagonal entry of $U_1$ may be taken to be of modulus
less than 1, by performing a further conjugation by
$\left( \begin{array}{cc}0 & -1 \\ 1 &0 \end{array} \right)\in  {\rm
SL}(2,R) $, if necessary. Thus any pair of type (AA) is equivalent to a
unique pair of the form (AA1) or (AA2) in the theorem.
\vskip .2cm
\noindent (AB) $U_2$ is equal to plus or minus the identity matrix, and thus
is unaffected by any conjugation. $U_1$ may be conjugated into a unique
diagonal form with the first diagonal entry of modulus less than 1 by a
matrix  $S\in  {\rm SL}(2,R) $, as in the previous case.
\vskip .2cm
\noindent (BB) Trivial, since both $U_1$ and $U_2$ are equal to plus or minus
the identity matrix.
\vskip .2cm
\noindent (BC) $U_2$ may be conjugated into the unique (Jordan) form
$\left( \begin{array}{cc} \epsilon_2 &1\\ 0&\epsilon_2\end{array} \right)$,
by $S'\in   {\rm GL}(2,R) $, where the first column of $S'$ is an
eigenvector of $U_2$ with eigenvalue $\epsilon_2=\pm 1$. If $\det S'>0$,
conjugating by $S=1/(\det S')^{1/2} S'\in  {\rm SL}(2,R) $ also puts $U_2$
into the same Jordan form. Otherwise, conjugating by
$$
S=1/(-\det S')^{1/2}
S' \left( \begin{array}{cc} -1 & 0\\ 0 & 1\end{array} \right) \in
{\rm SL}(2,R)
$$
puts $U_2$ into an alternative standard form
$\left( \begin{array}{cc} \epsilon_2 & -1\\ 0&\epsilon_2\end{array}
\right)$. Also, a direct calculation shows
$$
S^{-1} \left( \begin{array}{cc} \epsilon_2 &1\\ 0&\epsilon_2\end{array} \right)
S= \left( \begin{array}{cc} \epsilon_2 &-1\\ 0&\epsilon_2\end{array} \right)
\Longrightarrow S\notin {\rm SL}(2,R).
$$
Thus any pair of type (BC) is equivalent to a unique pair of the form (BC)
in the theorem.
\vskip .2cm
\noindent (BD)  $U_2$ may be conjugated uniquely into the real Jordan form
 $\left( \begin{array}{cc} a & b\\ -b&a\end{array} \right)$, with $a,b$
real and $b>0$, by $S'\in {\rm GL}(2,R) $. Since $a^2+b^2 = \det U_2=1$, one
may set $a=\cos \phi$, $b=-\sin \phi$, with $\pi<\phi<2\pi$. If $\det
S'>0$, conjugating by $S=1/(\det S')^{1/2} S'\in  {\rm SL}(2,R) $
also puts $U_2$ into the same real Jordan form. Otherwise, conjugating
by
$$
S=1/(-\det S')^{1/2} S'
\left( \begin{array}{cc} 0 & 1\\ 1 & 0\end{array} \right) \in
{\rm SL}(2,R)
$$
puts $U_2$ into the transposed Jordan form
$\left( \begin{array}{cc} \cos \phi & \sin \phi \\ -\sin \phi & \cos \phi
\end{array} \right)$, for $\pi<\phi<2\pi$, or equivalently, into the form
$ \left( \begin{array}{cc} \cos \phi & -\sin \phi \\ \sin \phi & \cos \phi
\end{array} \right)$, for $0<\phi<\pi$.  A direct calculation shows that
the matrices in Jordan form and its transposed form are not conjugate if
the conjugating matrix $S$ belongs to ${\rm SL}(2,R) $.  Thus any pair of
type (BD) is equivalent to a unique pair of the form (BD) in the theorem.
\vskip .2cm
\noindent (CC) $U_1$ may be conjugated into the unique Jordan form
$\left( \begin{array}{cc} \epsilon_1 &1\\ 0&\epsilon_1\end{array} \right)$,
with $\epsilon_1 = \pm 1$, by $S'\in   {\rm GL}(2,R) $. Let $v_1', \, v_2'$
denote the two columns of $S'$. Thus $v_1'$ is an eigenvector of $U_1$
corresponding to the eigenvalue $\epsilon_1$, and $v_2'$ satisfies
$
U_1v_2'=v_1' + \epsilon_1 v_2'
$.
Since $U_1$ and $U_2$ commute, $v_1'$ is also an eigenvector of $U_2$
(corresponding to the eigenvalue $\epsilon_2$). Now
\begin{eqnarray*}
U_1(U_2v_2'-\epsilon_2 v_2') &=& U_2U_1v_2' - \epsilon_2 U_1v_2' \\
&=& (U_2-\epsilon_2I)(v_1'+\epsilon_1v_2') \\
&=& \epsilon_1(U_2v_2'-\epsilon_2v_2')
\end{eqnarray*}
and therefore
$U_2v_2'-\epsilon_2v_2'=cv_1'$ for some $c\neq 0$. Let $\alpha \in
]0,2\pi[\setminus\left\{\frac{\pi}{2}, \pi, \frac{3\pi}{2}\right\}$ be
given by $\tan \alpha = c$ and ${\rm sgn} \cos \alpha ={\rm sgn} \det S'$.
Set $\tilde{v}_1 = (1/\cos \alpha) v_1'$, $\tilde{v}_2=v_2'$. Now
$U_1\tilde{v}_2= \cos \alpha\, \tilde{v}_1 + \epsilon_1\tilde{v}_2$ and
$U_2 \tilde{v}_2= \sin \alpha\, \tilde{v}_1 + \epsilon_2 \tilde{v}_2$, and
thus the matrix $\tilde{S}$ with columns $\tilde{v}_1$ and $\tilde{v}_2$,
and positive determinant, conjugates $U_1$ and $U_2$ into the form (CC)
above. The same holds for $S=(1/\det \tilde{S}^{1/2})\tilde{S}\in SL(2,R)$.
This form is unique, since suppose \[ S^{-1} \left( \begin{array}{cc}
\epsilon_1 &\cos \alpha\\ 0&\epsilon_1\end{array} \right) S = \left(
\begin{array}{cc} \epsilon_1 &\cos \beta\\ 0&\epsilon_1\end{array} \right)
\quad \quad S^{-1} \left( \begin{array}{cc} \epsilon_2 &\sin \alpha\\
0&\epsilon_2\end{array} \right) S = \left( \begin{array}{cc} \epsilon_2
&\sin \beta\\ 0&\epsilon_2\end{array} \right) \] for $S= \left(
\begin{array}{cc} a &b\\ c&d\end{array} \right)  \in SL(2,R)$. This implies
$c=0$, $\cos \alpha/\cos \beta = \sin \alpha/\sin \beta = a/d>0$, hence
$\alpha=\beta$. \vskip .2cm \noindent (DD) Regarded as complex matrices
$U_1$ and $U_2$ each have two conjugate complex eigenvalues of modulus 1,
say $e^{i\theta}$ and $e^{-i\theta}$ for $U_1$ and  $e^{i\phi}$ and
$e^{-i\phi}$ for $U_2$. Let ${u}_1$ be a joint eigenvector of $U_1$ and
$U_2$. Without loss of generality we may suppose that $U_1 {
u}_1=e^{i\theta} {u}_1$, $U_2 { u}_1=e^{i\phi} { u}_1$. Then ${
u}_2:=\bar{u}_1$ is a joint eigenvector corresponding to the eigenvalues
$e^{-i\theta}$ and $e^{-i\phi}$ respectively. Changing to a real basis
${v}_1={ u}_1 + {u}_2$, ${ v}_2=-i{u}_1 + i { u}_2$, $U_1$ and $U_2$ act as
follows: \begin{eqnarray*} U_1v_1&=&\cos \theta\, v_1 -\sin \theta\, v_2 \\
U_1v_2&=&\sin \theta\, v_1 +\cos \theta\, v_2 \\ U_2v_1&=&\cos \phi\, v_1
-\sin \phi\, v_2 \\ U_2v_2&=&\sin \phi\, v_1+ \cos \phi\, v_2.
\end{eqnarray*} Thus $U_1$ and $U_2$ are simultaneously conjugated into the
form DD) in the theorem by the matrix $S'\in {\rm GL}(2,R)$ which has
columns ${ v}_1$ and ${v}_2$. If $\det S'>0$, then conjugating by
$S=1/(\det S')^{1/2} S'\in {\rm SL}(2,R)$ also puts $U_1$ and $U_2$ into
the same form. If $\det S'<0$, then the matrix $\tilde{S}$ with columns
$-{v}_1$ and  ${v}_2$ has positive determinant, and conjugating with
$S=(1/\det\tilde{S})^{1/2} \tilde{S}\in {\rm SL}(2,R)$ puts $U_1$ and $U_2$
into the form DD) with the replacements $\theta \mapsto 2\pi - \theta$ and
$\phi \mapsto 2\pi - \phi$. Uniqueness of the form DD) follows from the
fact that the only conjugate pair of that form with the same spectrum
consists of the transposed matrices, but, as in the case (BD) above, a
matrix of this form and its transpose are not conjugate if the conjugating
matrix $S$ belongs to ${\rm SL}(2,R) $.  Thus any pair of type (DD) is
equivalent to a unique pair of the form (DD) in the theorem. \hfill\block

\vskip .2cm

\section{Discussion}

\begin{figure}[h]
\centerline{\psfig{figure=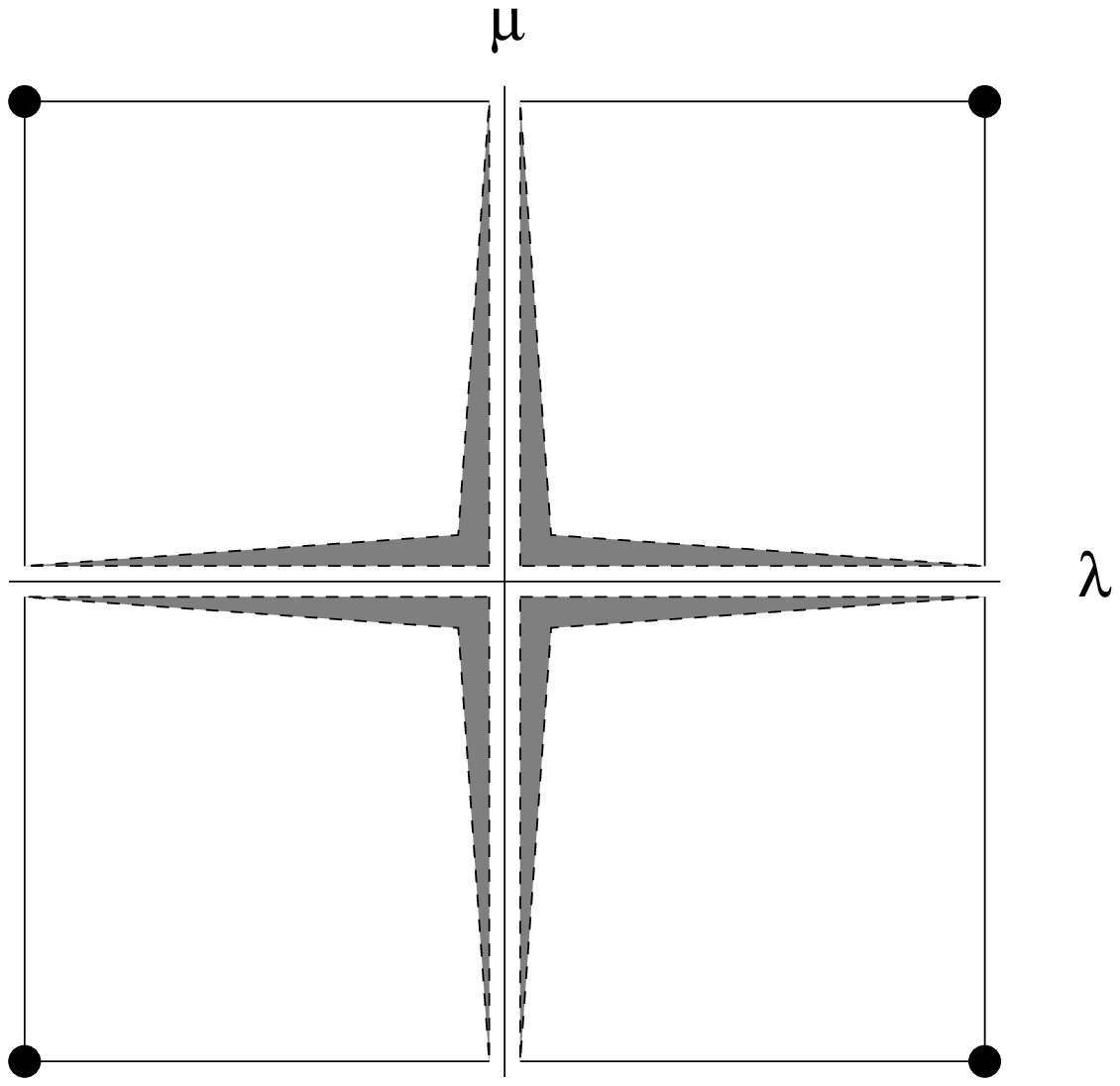,height=7cm,
}}
\caption{Subspace of A or B pairs}
\label{AB}
\end{figure}

\begin{figure}[h]
\centerline{\psfig{figure=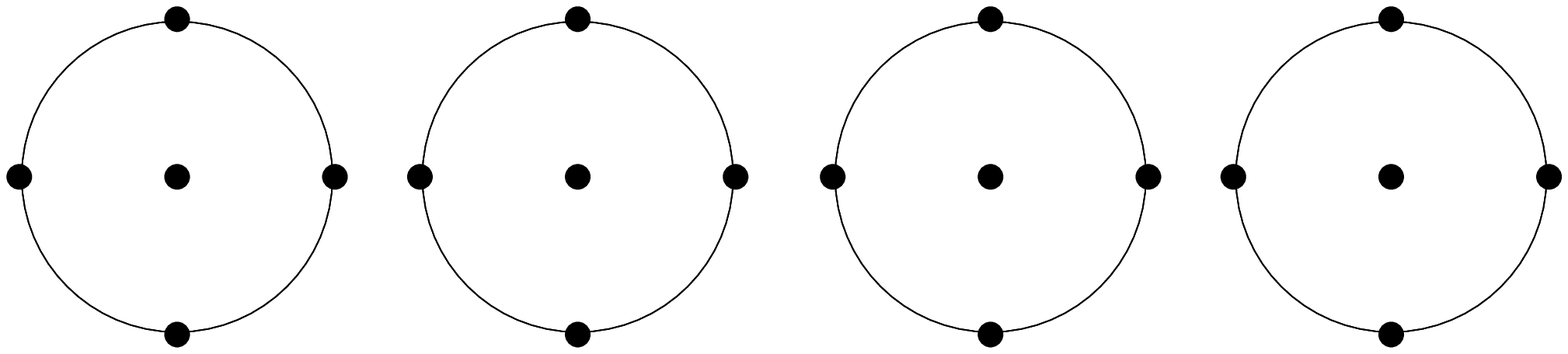,height=3cm,
}}
\caption{Subspace of B or C pairs}
\label{BC}
\end{figure}

\begin{figure}[h]
\centerline{\psfig{figure=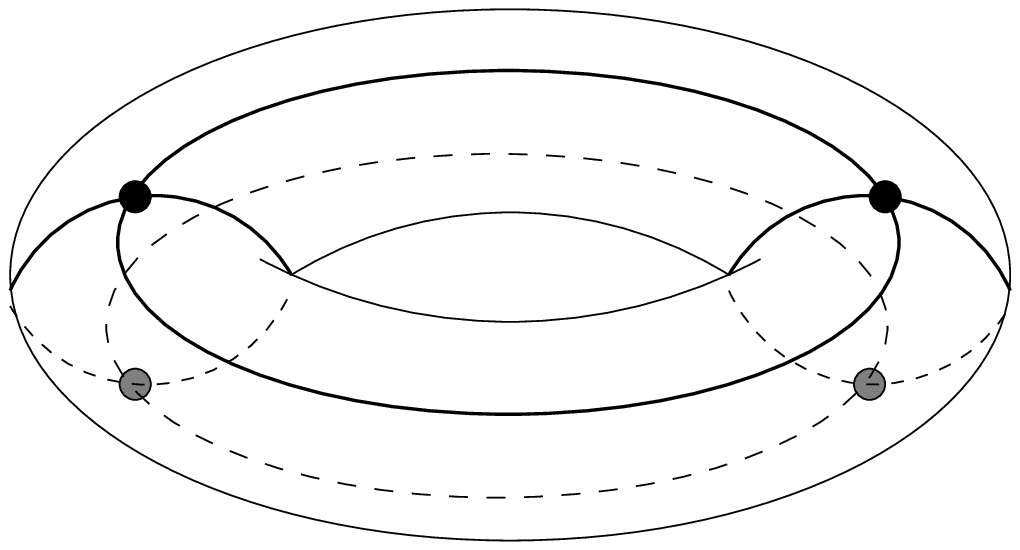,height=4cm,
}}
\caption{Subspace of B or D pairs}
\label{BD}
\end{figure}

\begin{figure}[h]
\centerline{\psfig{figure=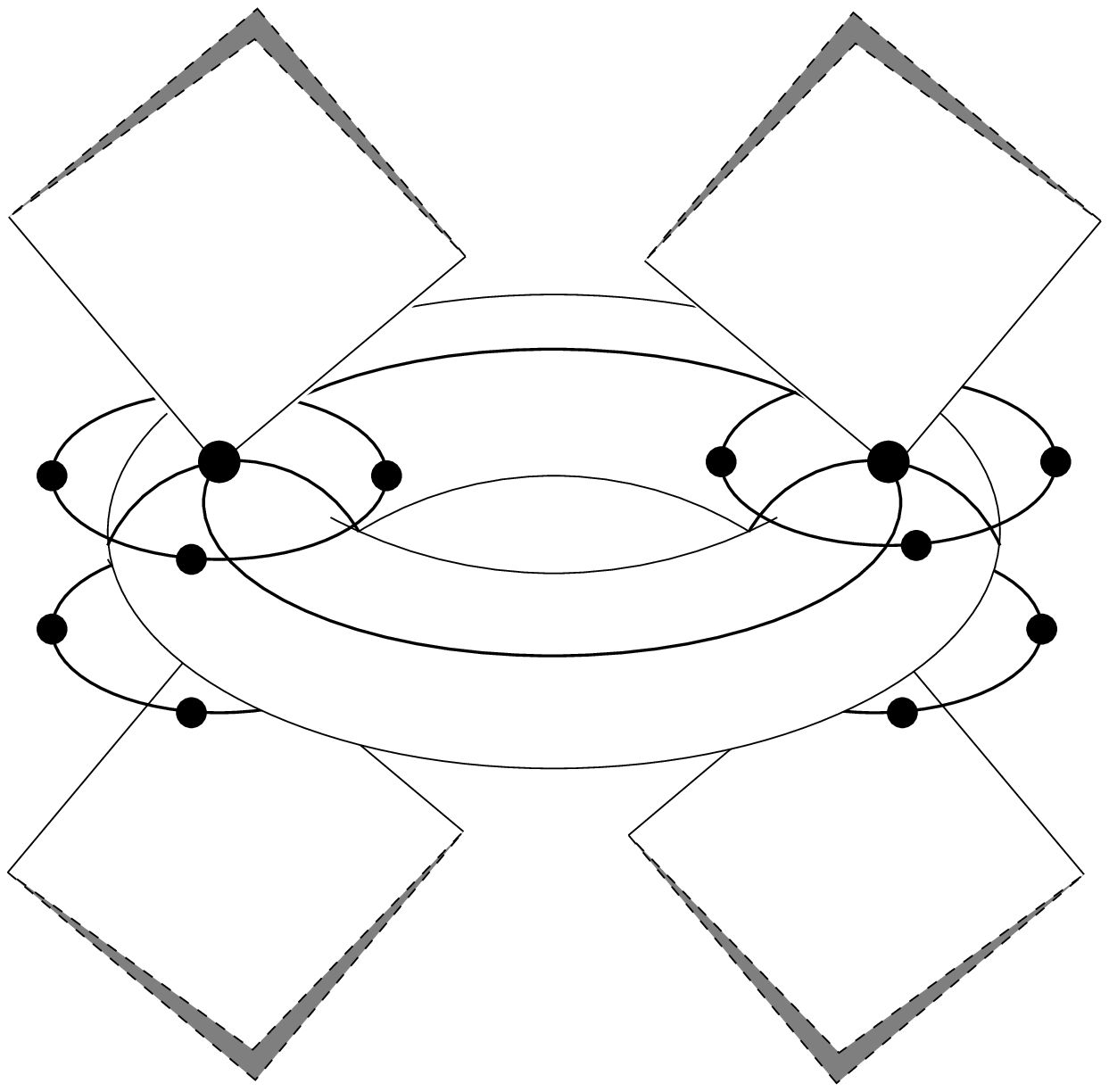,height=10cm,
}}
\caption{Overall view of the moduli space}
\label{overall}
\end{figure}

The theorem implies that we have an explicit parametrization of
the moduli space, which may be used to depict it. The subspace consisting
of pairs of matrices of type A or B (fig.~\ref{AB}) corresponds to a double
cover of the
region of the $(\lambda, \mu)$ plane $0<|\lambda|<1,\, 0<|\mu|<1$ (AA1 and
AA2 pairs), with the two sheets meeting along the lines $|\mu| = |\epsilon_2|=1, \,
0<|\lambda|<1$, and $|\lambda|=|\epsilon_1|=1, \,  0<|\mu|<1$ (AB and BA pairs), which in
turn meet at four corner points (BB pairs).
The subspace consisting of
pairs of type B or C (fig.~\ref{BC}) contains four points, coming
from the four BB pairs, and four associated circles,
 with each circle made up of four arcs (CC pairs) and four
intermediate points (BC and CB pairs).
The subspace consisting of pairs of
type B or D (fig.~\ref{BD}) is a torus parametrized by angles $\theta$ and
$\phi$ running over the full range $0$ to $2\pi$ made up of four open
regions with $\theta \neq 0, \pi$, $\phi \neq 0, \pi$ (DD pairs), eight
arcs with either $\theta =0, \pi$ or $\phi = 0, \pi$, but not both at the
same time (BD and DB pairs), and the four points $\theta =0, \pi$ and $\phi
= 0, \pi$ (BB pairs).  These subspaces are put together into a
single picture in figure~\ref{overall}. Represented in this way, the moduli
space resembles a cell complex (with open edges for the AA cells),
consisting of $2$-cells AA and DD attached to $1$-cells AB, BA, BD, DB,
which in turn are attached to $0$-cells BB, and separately $1$-cells CC
attached to $0$-cells BC and CB.

As a final point we wish to discuss the question of putting an appropriate
topology on the moduli space. The result of the theorem, and the depiction of
the moduli space in figure \ref{overall} which it gives rise to, suggest a
first natural choice, namely the topology induced by this representation as
a subspace of ${\bf R}^3$. In this topology the moduli space, although not
a manifold, is Hausdorff, and becomes a (non-compact) manifold after
excluding the four points corresponding to BB pairs. In slightly different
but related contexts the topology on the moduli space was found to be
non-Hausdorff. In \cite{ash:lew} Ashtekar and Lewandowski studied the
topology on the moduli space of all $SU(1,1)$ connections (not just flat
ones) modulo gauge transformations, using as a starting point a topology
on the space of all connections
compatible with the affine structure. Endowing the moduli space with
the quotient topology, gave rise to a non-Hausdorff topology. Louko
and Marolf in \cite{lou:mar} considered flat ${\rm ISO}(2,1)$ connections,
and used the quotient topology induced from the topology on ${\rm ISO}(2,1)
\times {\rm ISO}(2,1)$, also giving a non-Hausdorff topology on the
resulting moduli space. The comparison between these approaches leads one
to suspect that one should ``constrain before topologizing'' in order to
achieve the best-behaved topology.

We propose that the most appropriate topology to choose is that induced by
the parametrization of the theorem, but taking the eleven  sectors (AA1) to
(DD) to be mutually separated. Mathematically the separation between
matrices of type A (two one-dimensional eigenspaces), type B (one
two-dimensional eigenspace), type C (one one-dimensional eigenspace) and
type D (no eigenspaces) is justified by the difference between them in a
discrete spectral attribute (the number and dimension of their
eigenspaces). Physically one might argue that there is a significant
difference between a connection whose parallel transport around a
non-trivial cycle is trivial after every iteration ($\epsilon_i=1,\,
i=1,2$), or every two iterations ($\epsilon_i=-1,\,
i=1,2$), and one whose parallel transport converges or diverges
exponentially in the diagonal entries on iteration. In this topology each
sector is separately a manifold of dimension $0$, $1$ or $2$, with each
sector in turn consisting of separated components.

\vskip .5cm
\noindent {\bf Acknowledgements} R. Picken is grateful to N. Manojlovic and A.
Mikovic for useful discussions and to J. Baez for an interesting
conversation. This work was supported in part by the Istituto Nazionale di
Fisica Nucleare (INFN) of Italy Iniziativa Specifica FI41, the Italian Ministero
dell'Universit\`a e della Ricerca Scientifica e Tecnologica (MURST).
This work was supported by the Programa de Financiamento Plurianual and project
FCT/PRAXIS/2/2.1/FIS/286/94 of the
Funda\c{c}\~{a}o
para a Ci\^{e}ncia e a Tecnologia (FCT), and by the programme
{\em Programa Operacional
``Ci\^{e}ncia, Tecnologia, Inova\c{c}\~{a}o''} (POCTI) of the
FCT,
cofinanced
by the European Community fund FEDER.


\newpage

\vskip 3cm

\begin{thebibliography}{99}
\bibitem{ati:bot} Atiyah, M. F.; Bott, R.,``The Yang-Mills equations over Riemann
surfaces," Philos. Trans. Roy. Soc.
London Ser. A 308, 523--615 (1983).
\bibitem{gur:jef:hue:wei} Guruprasad, K.; Huebschmann, J.; Jeffrey, L.;
 Weinstein, A., ``Group
systems, groupoids, and moduli spaces of parabolic bundles,'' Duke Math. J. 89,
no. 2, 377--412 (1997).
\bibitem{gol}  Goldman, William M., ``The symplectic nature of fundamental groups of
surfaces," Adv. Math. 54, 200--225 (1984).
\bibitem{hit} Hitchin, N. J., ``The self-duality equations on a Riemann surface,'' Proc.
London Math. Soc. 55 (3), 59-126 (1987)
\bibitem{wit} Witten, E., ``2+1 dimensional quantum gravity as an exactly solvable
system,'' Nucl. Phys. B311, 46-78 (1988/89).
\bibitem{jeanette} Nelson J.E.; Regge T.; Zertuche F., ``Homotopy Groups and
$2+1$ dimensional Quantum De Sitter Gravity,'' Nucl. Phys.
B339, 516-532 (1990); Nelson J.E.; Regge T., ``$2+1$ Quantum Gravity,''
Phys. Lett. B272, 213-216 (1991); Carlip S.; Nelson J.E.,
``Comparative Quantizations of 2+1 Gravity,'' Phys. Rev. D51, 10, 5643-5653 (1995).
\bibitem{car} Carlip, S., ``Quantum gravity in $2+1$ dimensions,"
Cambridge Monographs on Mathematical Physics, Cambridge University Press,
Cambridge (1998).
\bibitem{NP1} Nelson, J. E.; Picken, R. F., ``Quantum holonomies in
$(2+1)$-dimensional gravity," Phys. Lett. B 471, 367--372 (2000).
\bibitem{NP2} Nelson, J. E.; Picken, R. F., ``Quantum matrix pairs," Lett.
Math. Phys. 52, 277--290 (2000).
\bibitem{CP} Caetano, A.; Picken, R. F.,
``An axiomatic definition of holonomy," Int. J. Math. 5, 835--848 (1994).
\bibitem{bar} Barrett, J. W., ``Holonomy and path structures in general
relativity and Yang-Mills theory," Internat. J. Theoret. Phys. 30,
1171--1215 (1991).
\bibitem{eza} Ezawa K., ``Reduced Phase Space of the First Order
Einstein Gravity on $\bf R \times \bf {T^2}$,'' Osaka preprint OU-HET-185 (1993),hep-th/9312151;
``Chern--Simons quantization of (2+1) anti-de Sitter gravity on a torus,''
Class. Quant. Grav. 12, 373-392 (1995); ``Classical and quantum evolutions of
the de Sitter and anti-de Sitter universes in (2+1) dimensions,'' Phys. Rev. D49,
5211-5226 (1994), Addendum-ibid D50, 2935-2938 (1995).
\bibitem{dwi:mat:nic} de Wit,
B.; Matschull, H.-J.; Nicolai, H., ``Physical states in $d=3,\;N=2$
supergravity,'' Phys. Lett. B 318, 115--121 (1993).
\bibitem{lou:mar} Louko, J.;
Marolf, D. M., ``The solution space of $2+1$ gravity on ${\bf R}\times T^2$
in Witten's connection formulation," Class. Quant. Grav. 11, 311--330
(1994).
\bibitem{lou} Louko, J., ``Witten's $2+1$ gravity on ${R} \times$
(Klein bottle)," Class. Quant. Grav. 12, 2441--2467 (1995).
\bibitem{fal:gaw} Falceto, F.; Gawedzki, K., ``Chern-Simons states at genus
one," Comm. Math. Phys. 159, 549--579 (1994).
\bibitem{ash:lew}
Ashtekar, A.; Lewandowski, J., ``Completeness of Wilson loop functionals on
the moduli space of ${\rm SL}(2,{C})$ and ${\rm SU}(1,1)$ connections,"
Class. Quant. Grav. 10, L69--L74 (1993).
\bibitem{mik:pic}
Mikovic, A.; Picken, R. F., ``Super Chern-Simons theory and flat super
connections on a torus," math-ph/0008006, to appear in Adv. Theor. Math. Phys.


\end{thebibliography}
\end{document}